# The polarized proton and antiproton beam project at U-70 accelerator

V.V. Abramov[1], I.L. Azhgirey[1], V.I. Garkusha[1,*], V.P. Kartashev[1], V.V. Mochalov[1,2], S.B. Nurushev[1,2], V.L. Rykov[2], P.A. Semenov[1,2], A.N. Vasiliev[1,2], V.N. Zapolsky[1], V.G. Zarucheisky[1]

[1] *Institute for High Energy Physics named by A.A. Logunov of National Research Centre "Kurchatov Institute", Protvino, Moscow region, 142281, Russia*
[2] *National Research Nuclear University MEPhI (Moscow Engineering Physics Institute), Moscow, 115409, Russia*

**Abstract**

The design and parameters of the polarized-beam facility at U-70 proton synchrotron of NRC "Kurchatov Institute" - IHEP are presented. The new beamline 24A will provide the polarized proton and antiproton beams for carrying out the rich physics program of the SPASCHARM experiment for comprehensive studies of spin phenomena in a wide spectrum of hadronic reactions in the energy range of 10-45 GeV.

## 1. Introduction

The planned upgrade of the U-70 accelerator [1] to a higher intensity of up to $3 \times 10^{13}$ protons per (9-10) s cycle opens new horizons for comprehensive studies in high energy physics at the accelerator complex of NRC "Kurchatov Institute" – IHEP. This requires, in turn, building the next-generation beamlines for delivering to the experiments the full spectrum of particles in the beams of the best quality. The higher proton beam intensity brings new challenges to the design of beamlines and requires a significant layout rearrangement in the existing U-70 experimental area. As a pilot project, two new beamlines, 24A & 24B, for the SPASCHARM [2] and VES [3] experiments are to be built at the areas currently occupied by some running experiments after their completion. These two beams will operate from a single external production target exposed to the slowly-extracted 50-60 GeV proton beam at the intensity of $\geq 10^{13}$ protons per cycle. Relying on an external target rather than currently used internal targets would essentially reduce the radiation load to the equipment in the U-70 ring. This also provides more flexibility for designing the higher intensity and better quality beams and significantly improve an utilization efficiency of protons accelerated in U-70.

This paper is focused on the design and properties of the beamline 24A which will deliver a whole spectrum of charged-particle beams to the SPASCHARM experiment. The SPASCHARM's physics program requires hadronic beams of various species as well as electrons and/or positrons for detector calibrations. But predominantly it concentrates on systematic and comprehensive studies of spin phenomena in exclusive and inclusive hadronic reactions in the U-70 energy range. Therefore, the core SPASCHARM program [2] relies, first of all, on the study of interactions of high energy polarized protons and antiprotons selected and transported to the experiment by the beamline 24A, as well as on the use of polarized target.

The polarized proton and antiproton beams for the SPASCHARM experiment are to be created by the method suggested by O. Overseth and J. Sandweiss [4]. So far, this method has been twice successfully realized in practice. First, the 185 GeV/c proton and antiproton polarized beam facility has been built and operated at an 800 GeV accelerator of Fermilab for the E704 experiment [5]. Later on, the existing multipurpose beamline at U-70 accelerator has been modified in order to form the 40 GeV/c polarized proton beam for the FODS experiment [6].

While the main subject of this paper is the discussion specifically of the beamline 24A, it is preceded by the description of the target area, where the slowly extracted proton beam from U-70 is steered onto the external target, and then the secondary particle flux is split and directed toward the simultaneously operated beamlines 24A & 24B.

## 2. Target area for the 24A & 24B beamlines

The schematic view of the target area is shown in **Fig. 1**. With the dipole magnets MT1 & MT2, the primary proton beam from U-70 is directed through the target T toward the center point of the magnet MT3 at certain angle $\varphi$ in horizontal plane relative to the MT1-MT3's center line. Then, the third dipole magnet MT3 distributes the fractions of the secondary particle flux, produced in the target, between the beamlines 24A & 24B. Two parameters: the angle $\varphi$ and the MT3 bending power are adjusted so as the particles of desirable charges and momenta are directed toward the beamlines 24A & 24B. Such "three-magnet systems" have been successfully used at CERN SPS experimental areas for operating two or three beams from a single production target [7].

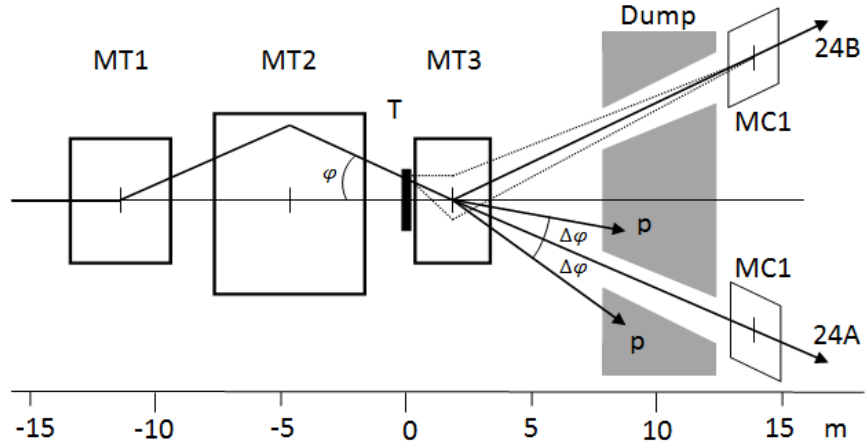

**Fig. 1.** Schematic top view of the target area for the beamlines 24A & 24B. T is for production target; MT is for dipole magnets; MC1 is for magnet-correctors; Dump is the 5 m thick steel beam absorber. In the shown configuration, neutral secondaries go straight toward the 24A beamline, while charged particles of a desirable sign are deflected toward beamline 24B. The dashed lines show some trajectories of charged particles emitted from the target at nonzero angles.

The dipole magnets SP-129 & SP-7[1] from the equipment pool of the U-70 experimental complex will be used for MT1 & MT2. The angle $\varphi$ is limited to $|\varphi| \leq \varphi_{max} = 27$ mrad by the parameters of the dipole magnet SP-7. The initial sections of the beamlines 24A & 24B are aligned along the straight lines through the MT3 center at the angles $\pm\varphi_{max}$. On the one hand, the large angle between these directions, equal to $2\varphi_{max}$, gives more space for placing the first quadrupole lenses as close to the target as it possible. On the other hand, with such a layout, it is still possible to direct towards either beamline a flux of neutral particles produced in the target at zero angle relative to the direction of the incident primary protons.

---

[1] SP-129 parameters: length $L = 4$ m, useful aperture $H \times V = 330 \times 100$ mm$^2$, maximum field $B_{max} = 1.8$ T; SP-7 parameters: $L = 6$ m, $H \times V = 500 \times 200$ mm$^2$; $B_{max} = 1.8$ T.

As the basic element in the target area the MT3 dipole magnet operates as a distributor of charged particles, thereby defining the charges and momenta for the both beamlines. At the same time, it serves as a sweeping magnet for "neutral" beams, deflecting all charged particles out of the beamline, which is used to create the beams of polarized protons (antiprotons) from $\Lambda(\overline{\Lambda})$-decays as well as of electron or positron beams from $\gamma$-conversions [8]. As a result, MT3 parameters have been chosen as a compromise between contradictory requirements to its length at the highest value of the magnetic field: $L = 2.6$ m, useful aperture $H \times V = 140 \times 56$ mm$^2$, $B_{max} = 1.9$ T.

The production target, which is located at the front end of the MT3 magnet, is an aluminum plate, 400 mm long, 3 mm high, and 100 mm wide. The extended horizontal width of the target makes it unnecessary to move the target when the angle $\varphi$ varies within the limits $|\varphi| \leq \varphi_{max}$.

The target area layout is symmetric relative to the initial direction of the primary proton beam. This makes the beamlines 24A & 24B absolutely equivalent in terms of the capabilities for directing any available particle flux to the acceptance of either beamline. The relations between parameters of two beams are as follows:

- With the angle $\varphi = \pm\varphi_{max}$, the secondary neutral particles produced at zero angles go straight in the direction of one of the beamlines. At the same time, the beam of positive or negative particles with momentum in the range from 16 to 28 GeV/c is deflected towards the other beamline. The lower momentum limit is set by the requirement for non-interacted primary protons to be deflected by MT3 away from the angular range $\pm(\varphi_{max} \pm \Delta\varphi)$, and steered to the absorber. The highest available momentum is limited by the maximum deflection power of the magnet MT3.

- With the angle $|\varphi| < \varphi_{max}$, the secondary particles of opposite charges are selected for the beamlines 24A & 24B. In the most interesting case of utilizing secondary particles produced at zero angles, the momenta $p_A$ and $p_B$ are bound by the relation: $q_A p_A(\varphi_{max} + \varphi) + q_B p_B(\varphi_{max} - \varphi) = 0$ where $q_A = -q_B = \pm 1$ are the particle charges. As in the previous case, the requirement of correct absorption of non-interacted primary protons sets certain limitations on momenta for the available secondary particles produced at zero angles.

In both cases, the magnet-correctors MC1 with the deflection power of up to 0.3 T×m in the horizontal plane extend the beamline acceptances to charged secondary particles produced in the target at non-zero angles.

The MT3 magnet is located just downstream the target struck by a high intensity primary proton beam. Hence, it is irradiated by a high intensity flux of secondary particles emerging from the target. This sets strong requirements to its long term radiation hardness, resulting in certain limitations on the MT3 design and available technologies. The most vulnerable part of the magnet under high radiation is the insulation of the exciting coil. In order to reduce the radiation load, the MT3 coil was moved upward above the beamline plane by ~35 cm from the direct view of the target behind the upper pole steel, as shown in **Fig. 2**. Such a solution [9] allows to save on investment in the technologies for fabricating the coils with, e.g., asbestos-cement or MgO radiation-resistant insulations [10,11] but rely instead on a coil manufactured, using more conventional technology with vacuum impregnation by epoxide compound.

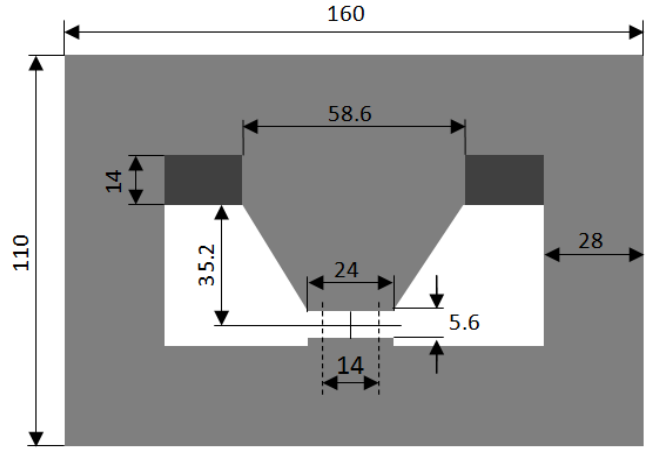

**Fig. 2.** Cross section of the MT3 magnet. All dimensions are in cm.

The simulations of radiation load, using the MARS computer code [12], have shown that, in the magnet design described above, the estimated life-time of the coil (the time required to accumulate a radiation dose of 10 MGy [10]) is about 300 days at the intensity of primary protons of $2 \times 10^{13}$ per cycle. With the additional steel plates, protecting the protruding parts of the coil, and the concrete filling of the space between the upper pole and the yoke, the life-time would be extended to ~2600 days.

## 3. Optical scheme of the polarized proton (antiproton) beamline 24A

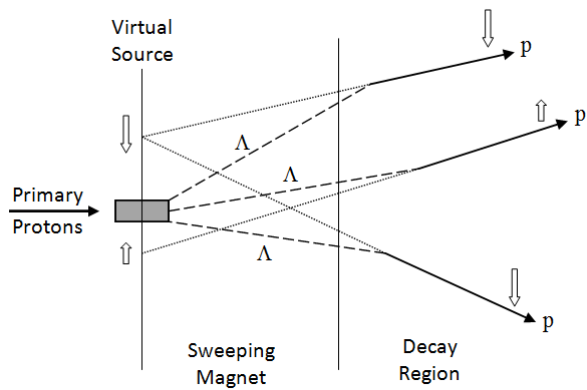

**Fig. 3.** A simplified scheme of correlation between proton's transverse polarization and its track's crossing position back to the plane of the virtual source in $\Lambda \rightarrow p\pi^-$ decay [5].

The method [4-6] for creating polarized proton or antiproton tertiary beams, which exploits the parity-violating decays of $\Lambda$-hyperons, is schematically illustrated in **Fig. 3**. In the $\Lambda$-hyperon rest frames, protons from $\Lambda \rightarrow p\pi^-$ decays [2] are longitudinally polarized with the helicity equal to the decay constant $\alpha$. The experimentally measured $\alpha=0.642\pm0.013$ [13]. After the Lorentz boost into the laboratory frame, the longitudinal polarization is partially converted into a transverse polarization relative to the proton momentum. This polarization is larger for larger decay angles. In other terms, the transverse polarization correlates with the positions of crossing points of proton trajectories, traced back to the transverse virtual-source plane through the center of the production target.

---

[2] The kinematics of $\Lambda \rightarrow p\pi^-$ and $\overline{\Lambda} \rightarrow \overline{p}\pi^+$ decays is identical except of the opposite signs of charges and polarizations of decay protons and antiprotons.

The transverse polarization, averaged over all decay protons, is obviously zero. But one can select samples of a nonzero mean polarization by sorting out proton trajectories over their decay angles or, which turned out to be more practical [5,6], over the crossing points of trajectories traced back to the virtual source. The maximum decay angle in the laboratory frame and, consequently, the full size of the virtual source depends on the proton momentum: the angle and size decrease as the momentum increases. It worth underlining that the full size of the virtual source would be smaller with the shorter sweeping magnet (MT3 in **Fig. 1**).

Actually, the position-polarization correlations are smeared with the longitudinal and transverse positions of Λ-hyperon production points in the target by a primary beam of a finite transverse size, with the distance of an actual Λ-decay from the virtual-source plane, as well as with the momentum spread of the decay protons captured into the beam. The capture into the beam of some number of protons from Λ-decays inside the sweeping magnet just before its exit causes additional smearing of the position-polarization correlation in the MT3 bending plane. This is the main reason why the sorting out of trajectories over the polarization usually takes place along only one axis in the plane orthogonal to the beamline axis [5,6] which is the vertical axis $y$ in our case.

The primary requirement for the beam transport system 24A is its capability to form and deliver to the experiment the tertiary beams of decay protons and antiprotons of desired momenta with the highest achievable intensity and low background and at minimal losses of polarization, as well as providing means for sorting out beam rays by polarization. The beam optics described below is optimized so as to create the beams of the best quality in full compliance with the basic requirements [5,14] for such systems, and taking into account the constraints associated with the availability of space in the U-70 experimental area.

The main optical scheme of the beamline 24A, developed to transport beams of polarized protons and antiprotons from $Λ(\overline{Λ})$-decays [15], is shown in **Fig. 4**. The optical system consists of two mirror sections separated by intermediate focuses in both, the horizontal and vertical planes, where the sorting and selection of the beam trajectories with respect to momentum and polarization takes place. The final focuses are at the experiment target.

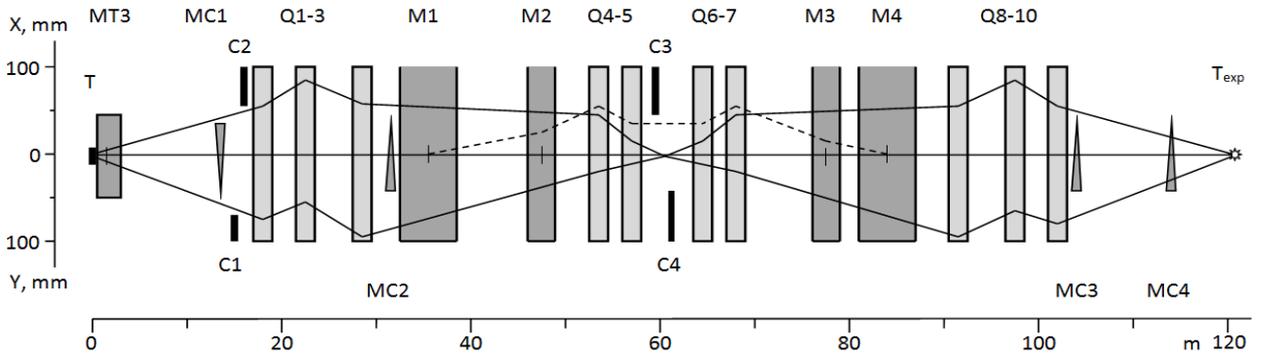

**Fig. 4.** The main optical scheme of the beamline 24A. Q is for quadrupole lenses; M is for dipole magnets; C is for collimators; MC is for magnet-correctors; T and $T_{exp}$ are the production and the experiment targets, respectively. Solid lines show beam focusing in both transverse planes. The dashed line shows the dispersion of the beam in the horizontal plane for $Δp/p_0=5\%$.

An identity transformation matrix of optics in the vertical plane from the virtual source to the final focus is exploited. This ensures that, in the linear approximation, the vertical component of polarization $\zeta_y$ of a beam ray does not change [14]. In other words, in the linear order, there is no of the beam depolarization along the path from $\Lambda(\overline{\Lambda})$-hyperons decay points down to the target of the experiment. The mirror symmetry of two sections in the chosen optical system leads to the transformation in the horizontal plane to be also close to the identity one.

The momentum analysis takes place in the horizontal plane in the intermediate focus, using the collimator C3, where the momentum dispersion of 7.1 mm per one percent of $\Delta p/p_0$ is created by the dipole magnets M1 & M2. Then, it is compensated downstream by the dipole magnets M3 & M4 through quadrupole lenses Q4-Q7. The magnification coefficient $x/x_0$ of the transformation matrix from the production target to the intermediate focus is equal to ~0.5. The momentum spread in the beam propagating, downstream the collimator C3, depends on its opening. All four dipole magnets M1-M4 deflect the beam in the horizontal plane toward one side to a total angle of 147 mrad.

Sorting out over the vertical polarization component $\zeta_y$ and selecting polarized samples from the full unpolarized beam takes place in the vertical intermediate focus, where the image of the virtual source is magnified in order to maintain near the same vertical size irrespectively of the momentum. Thus, the measurement of the position $y$ for each proton (antiproton) trajectory in the intermediate focus is equivalent to measuring the mean polarization associated with each particular trajectory. This is the essence of the beam-tagging method [5,16], using a number of fast beam hodoscopes for tracing the beam particles and measuring their momenta. The beam-tagging hodoscopes will also be used for tuning the beam.

The other approach for selecting the polarized beam samples is by cutting out a fraction of the beam, using a vertical collimator C4 along with the vertical magnet-correctors MC2-MC4 [6]. In this case the horizontal slit of the collimator C4 is always centered vertically at the beamline axis, and the desirable part of the beam is moved up or down and guided into the slit by the MC2 magnet-corrector. MC2 is positioned in the optical scheme so as to not change the direction of the beam rays in the intermediate focus when the entire beam is moved up or down. Magnet-correctors MC3 & MC4 then are used to steer the selected beam sample back to the experiment's axis. By reversing the currents in all three magnet-correctors, the other beam sample with the mean polarization of the opposite sign is sent to the experiment target. Such an inversion of polarization can be performed quickly within every accelerator cycle, which is important for reducing systematic errors in the measurements of the spin asymmetries.

In order to carry out the SPASCHARM studies with longitudinally polarized protons (antiprotons), there are plans to install in the final stretch of the beamline (downstream of the triplet Q8-Q10) a ~8 m long superconducting helical spin-rotator [17], which can also be used for periodic inversions of transverse polarization at the experiment target.

The usual equipment of the U-70 experimental complex will be used in the beamline 24A: quadrupole lenses 20K200 (diameter of aperture 200 mm, magnetic field gradient up to 13 T/m, length 2 m); dipole magnets SP-7 & SP-12 (useful aperture $H \times V = 500 \times 200$ mm², magnetic field up to 1.8 T, lengths 6 & 3 m, respectively); collimators (maximum slit opening ±75 mm, length 0.75 m).

The beamline angular acceptance is limited by the collimators C1 & C2. The acceptance limits of ±2.8 mrad in vertical plane and ±3.2 mrad in horizontal plane are chosen so as to propagate the neutral particle flux downstream the collimators, predominantly neutrons, fully confined within the apertures of the vacuum chambers of quadrupole lenses Q1-Q3 and dipole magnet M1, thereby effectively eliminating the background from interactions of the neutrals in these elements. Neutral particles that pass through the collimators C1 & C2 are eventually dumped into the second absorber (not shown in **Fig. 4**), located between the dipole magnets M1 & M2 at a distance of 6 m from the center of M1. The deflection angle in M1 for a charged beam is equal to 52 mrad.

The maximum beam momentum in the main optical scheme is 45 GeV/c. Other characteristics of polarized proton and antiproton beams from $\Lambda(\overline{\Lambda})$-decays, such as available momentum spread, beam dimensions, etc., essentially depend on the virtual source size, which, in turn, varies with momentum. For example, the size of the full beam at the target of the experiment, which is transferred one-to-one from the virtual source by means of the identity optical transformation, becomes smaller as the momentum increases and vice versa.

With the first (Q1-Q3) and last (Q8-Q10) triplets of quadrupole lenses powered on as doublets, charged particles of momenta up to 60 GeV/c could be transported from the production target to the SPASCHARM experiment. This also includes the option for delivering to the experiment the primary protons at reduced intensity, extracted from the U-70 by a bent crystal [18]. For the beams of charged secondaries from the production target, the minimum attainable full momentum-band width $\Delta p/p_0$ would be quite small, just ±1.2%, because of the smaller and momentum-independent transverse size of the particle source, in comparison to the size of the virtual source of protons (antiprotons) from $\Lambda(\overline{\Lambda})$-decays.

## 4. Beam parameters at the intermediate focus

In order to have approximately the same resolution with respect to the vertical component of polarization $\zeta_y$ for different beam momenta, using the same beam-tagging hodoscopes, it is desirable to keep the vertical size of the virtual-source image in the intermediate focus independent of momentum. This is achieved by adjusting the magnification coefficient $y/y_0$ of the transformation matrix from the production target to the intermediate focus. The main parameters of the polarized proton beams in the intermediate focus are presented in **Table 1**. The values of $\sigma_x$ also include the contributions of momentum dispersion. The minimum momentum spread is limited by the finite horizontal size of the virtual-source image at the momentum collimator (C3 in **Fig. 4**).

**Table 1.** The simulated parameters of proton beams in the intermediate focus. The numbers for the beam dimensions are given for minimum and maximum (in brackets) momentum spread, transmittable through the beamline. Here and everywhere throughout the paper, the notation 'σ' is used for the RMS sizes of the respective distributions.

| Beam momentum setting $p_0$ (GeV/c) | 15 | 30 | 45 |
|---|---|---|---|
| Vertical size $\sigma_y$ of the virtual source (mm) | 14.5 | 10.1 | 8.5 |
| Vertical magnification coefficient $y/y_0$ for $p_0$ | 1.4 | 2.0 | 2.4 |
| Momentum spread $\sigma_p/p_0$ (%) | 2.0 (4.5) | 1.4 (4.4) | 1.2 (4.2) |
| Vertical beam size $\sigma_y$ in the intermediate focus ( mm) | 20.6 (22.1) | 20.9 (22.0) | 21.3 (21.7) |
| Horizontal beam size $\sigma_x$ in the intermediate focus (mm) | 10.0 (32.2) | 7.7 (37.0) | 6.6 (35.0) |

The simulated correlations between the ray vertical position $y$ in the intermediate focus and the average per a ray polarization $\zeta_y$ at the experiment target are shown in **Fig. 5** for the 45 GeV/c proton beam. For a wider momentum band, the values of $\zeta_y$ for the same $y$ are somewhat lower due to more smearing of the correlations due to larger effect of chromatic aberrations.

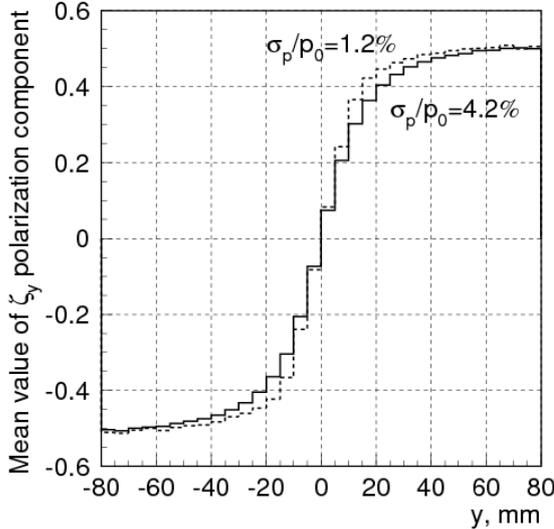

**Fig. 5.** Correlation between the average vertical component of polarization $\zeta_y$ at the experiment target and the trajectory vertical position $y$ at the intermediate focus for the 45 GeV/c proton beam.

All the beam simulation results throughout this paper have been obtained, using the Monte-Carlo computer code DECAY TURTLE [19], modified in order to accommodate the propagation of polarization vectors in magnetic optical elements for protons (antiprotons) from $\Lambda(\overline{\Lambda})$-decays. The yields of $\Lambda(\overline{\Lambda})$-hyperons produced in the production target by 60 GeV primary protons have been simulated, using the phenomenological formulae from Ref. [20]. The estimates for the beam intensities are normalized to the exposure of the production target to $10^{13}$ protons.

## 5. Parameters of polarized proton beams in the final focus at the experiment target

The results of Monte-Carlo simulations for the parameters of polarized proton beams at the SPASCHARM target are presented in **Table 2** for three beam momentum settings. These include the spatial dimensions $\sigma_x$,$\sigma_y$, the angular divergences $\sigma_{x'}$,$\sigma_{y'}$, and the beam intensities $I_{p(\Lambda)}$, as well as the estimates for $\pi^+$-meson background $I_{\pi^+(K_s^0)}$ from $K_s^0 \to \pi^+\pi^-$ decays, both per $10^{13}$ of primary protons on target (pot). These decays constitute the main background source provided that the magnet MT3 (**Fig. 1**) effectively cleans up the beam from positive secondary particles, generated directly in the production target. For polarized proton beams of momenta $\geq$30 GeV/c, this could be well done if the magnetic field in the MT3 was near a maximum of 1.9 T. This corresponds to the selection for the beam 24B of secondary particles, produced at zero angles, with momenta close to a maximum of 28 GeV/c. With the help of the magnet-corrector MC1 the available momentum range in the beamline 24B can be expanded by steering the secondary particles produced at nonzero angles to the beamline 24B.

**Table 2.** The simulated characteristics of proton beams at the SPASCHAM target.

| $p_0$ (GeV/c) | 15 | | 30 | | 45 | |
|---|---|---|---|---|---|---|
| $\sigma_p/p_0$ (%) | 2.0 | 4.5 | 1.4 | 4.4 | 1.2 | 4.2 |
| $\sigma_x \times \sigma_y$ (mm) | $17 \times 14$ | $19 \times 16$ | $14 \times 10$ | $17 \times 11$ | $11 \times 8.7$ | $16 \times 9.0$ |
| $\sigma_{x'} \times \sigma_{y'}$ (mrad) | $1.4 \times 1.5$ | $1.3 \times 1.5$ | $1.5 \times 1.8$ | $1.3 \times 1.8$ | $1.4 \times 1.7$ | $1.3 \times 1.7$ |
| $I_{p(\Lambda)}$ / $10^{13}$ pot | $3.5 \times 10^6$ | $9.2 \times 10^6$ | $2.1 \times 10^7$ | $7.8 \times 10^7$ | $1.5 \times 10^7$ | $6.8 \times 10^7$ |
| $I_{\pi^+(K_s^0)}$ / $10^{13}$ pot | $3.8 \times 10^5$ | $1.1 \times 10^6$ | $3.5 \times 10^5$ | $1.4 \times 10^6$ | $1.3 \times 10^4$ | $7.6 \times 10^4$ |

Because of the identity transformation matrix in both planes (see **Sec. 3**), the vertical spatial dimension of the beam at the experiment target for narrow momentum spread follows the sizes of the virtual source. It is larger at low momenta, and decreases as the momentum increases. There is also some contribution of chromatic aberration in both dimensions, which is particularly noticeable for the larger momentum spreads. As an example, the spatial beam profiles at the experiment target are shown in **Fig. 6** for the 45 GeV/c proton beam with $\sigma_p/p_0 = 1.2\%$ and for the 15 GeV/c proton beam with $\sigma_p/p_0 = 4.5\%$.

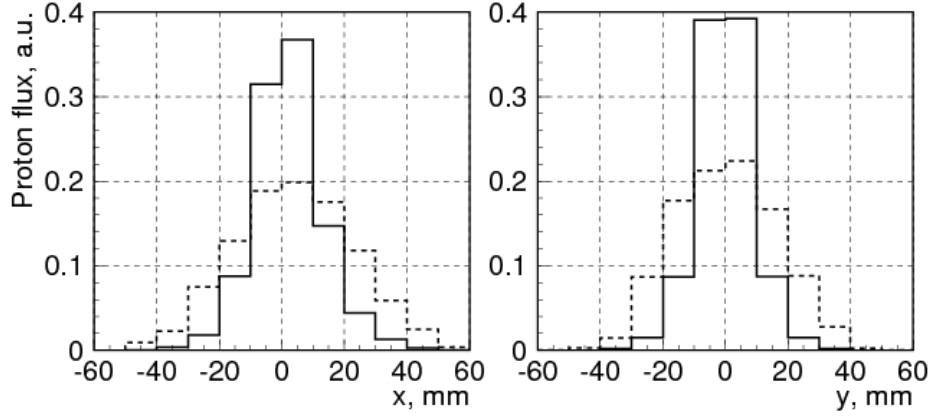

**Fig. 6.** Spatial profiles at the experiment target of the 45 GeV/c proton beam with $\sigma_p/p_0 = 1.2\%$ (solid lines) and the 15 GeV/c proton beam with $\sigma_p/p_0 = 4.5\%$ (dashed lines).

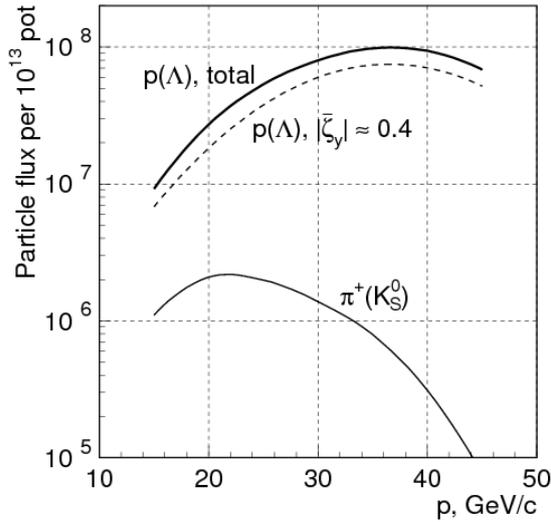

**Fig. 7.** The proton beam intensity at the experiment target as a function of the beam momentum for the maximum transmittable $\Delta p/p_0$, along with the background from $K_s^0$ decays. The dashed line shows the summary intensity of two samples with the opposite average polarization ±40%.

The intensity of polarized protons as a function of the beam momentum is shown in **Fig. 7** for the maximum transmittable $\Delta p/p_0$ along with the background from $K_s^0 \rightarrow \pi^+\pi^-$ decays. As it follows from **Table 2** and **Fig. 7**, the background from $K_s^0$-decays is rather small, just about 10-12% at 15 GeV/c, and its relative contribution rapidly drops down as the beam momentum increases.

Beam samples with nonzero average vertical polarization $\bar{\zeta}_y$ are built from the unpolarized full beam by selecting trajectories with their vertical position in the intermediate focus at $|y| > y_{cut}$. Polarizations $\bar{\xi}_y$ for the samples at $y < -y_{cut}$ and $y > y_{cut}$ are equal, but of opposite signs.

If the collimator C4 along with the magnet-correctors MC2-MC4 were used to select polarized samples, as described in **Sec. 3**, a portion of the beam, propagating downstream through the collimator C4, would be defined by the deflection power of the corrector MC2. As an example, the dependences of the parameters of a polarized proton beams at the target of the experiment on the MC2 deflection power are presented in **Table 3** for $p = 45$ GeV/c, $\sigma_p/p_0 = 1.2\%$, and with the fixed to ±15 mm opening of the C4 slit.

**Table 3.** Parameters of polarized proton beams at the experiment target as functions of the deflection power of the magnet-corrector MC2 for $p = 45$ GeV/c, $\sigma_p/p_0 = 1.2\%$, and C4 opening equal to $\pm 15$ mm. The mean polarization values $\overline{\zeta_y}$ are given along with the RMS width of $\zeta_y$-distributions over the selected samples.

| $(BL)_{MC2}$, T×m | 0.06 | 0.08 | 0.10 | 0.12 |
|---|---|---|---|---|
| $\overline{\zeta_y}$ | | 0.32±0.19 | 0.39±0.15 | 0.44±0.14 | 0.47±0.13 |
| $I_{p(A)}$ / $10^{13}$ pot | | $6.5\times10^6$ | $5.2\times10^6$ | $3.8\times10^6$ | $2.5\times10^6$ |

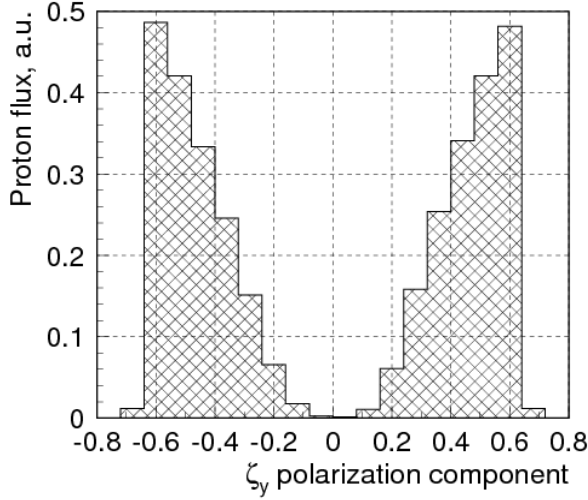

**Fig. 8.** The proton distributions with respect to the polarization $\zeta_y$ for the beam samples at MC2 deflections powers of $\pm0.12$ T×m.

In the horizontal plane, all the parameters of the selected polarized samples are the same as those shown in **Table 2** and **Fig. 6** for the full beam. The vertical spatial dimension of the selected polarized fractions of the total beam at the experiment target are expectedly smaller with $\sigma_y$ just about 3.2 мм. The mean polarization $\overline{\zeta_y}$ changes its sign with the current reversal in MC2-MC4. Typical distributions of protons over the polarization $\zeta_y$ in the beam samples with nonzero $\overline{\zeta_y}$ are shown in **Fig. 8**. All other parameters are the same as in **Table 3**.

## 6.  Parameters of polarized antiproton beams

As it has been already mentioned in **Sec. 3**, the kinematics of $\overline{A} \to \overline{p}\pi^+$ decay which serves as a source of polarized antiprotons is identical to the one of $A \to p\pi^-$ decay. Therefore, all characteristics of antiproton beams are the same as presented in **Tables 1-3** and **Fig. 5, 8,** with exception of intensities, background and the opposite signs of polarization. The antiproton intensities are much lower than for protons due to the significantly smaller production cross section for $\overline{A}$-hyperons compared to $A$-hyperons. For example, the intensity of the 15 GeV/c antiproton beam is lower by a factor of ~20 than of the proton beam of the same momentum, and the differences are even greater at higher momenta. The same cause lies behind the entry into play here of one more background source: $\pi^-$-mesons from $A \to p\pi^-$ decays (for protons, the $\pi^+$ background from $\overline{A} \to \overline{p}\pi^+$ is negligibly small). The maximum achievable antiproton beam intensities at the experiment target along with the backgrounds are shown in **Fig. 9** as functions of the beam momentum.

Apparently, the background from $A \to p\pi^-$ at low momenta is very high, but fortunately, it has a fairly sharp cut off at ~15 GeV/c due to specific properties of $A$-decay kinematics. It follows that the lowest $\pi^-/\overline{p}$ ratio of only ~3 is reached at about 16 GeV/c, where the intensity of antiproton beam would be ~$4\times10^5$ particles per accelerator cycle. The spatial distributions of the 16 GeV/c antiproton beam at the experiment target for the maximum transmittable $\Delta p/p_0$ are close to those for 15 GeV/c proton beam in **Fig. 6**.

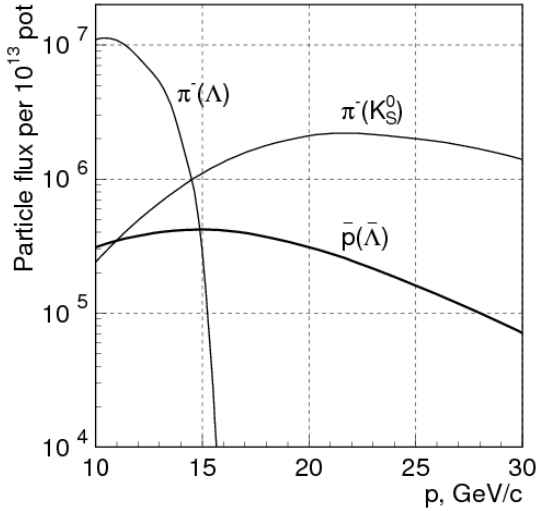

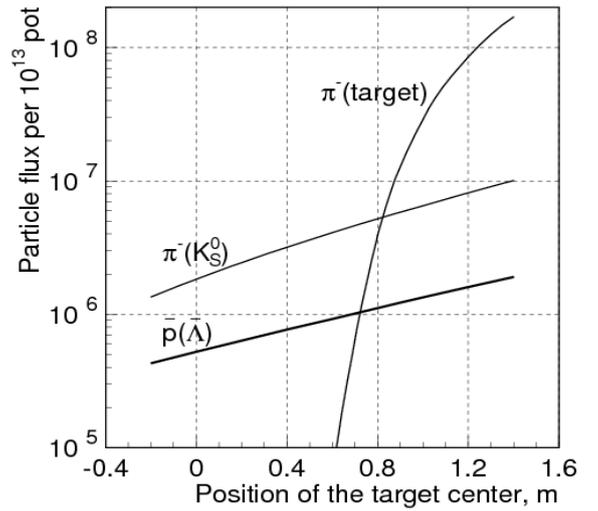

**Fig. 9.** The antiproton beam intensity and backgrounds at the experiment target as functions of the beam momentum for the maximum transmittable $\Delta p/p_0$.

**Fig. 10.** Dependences of the antiproton beam intensity and of the $\boldsymbol{\pi^-}$-backgrounds on the position of the production target with respect to the upstream edge of the MT3 magnet for the 16 GeV/c beam with the maximum transmittable $\Delta p/p_0$. The position of the main target center corresponds to −0.2 m.

In the beamline configurations described in **Sec. 2, 3**, this is, in fact, close to the maximum achievable intensity of antiprotons per $10^{13}$ primary protons hitting the production target. Still, there is an option to increase antiproton intensity by replacing the main production target with the other one installed inside the sweeping magnet MT3. If this target is positioned at some depth into the sweeping magnet MT3, the smaller fraction of produced $\overline{\Lambda}$-hyperons would be lost to useless decays within the MT3. However, such a modification would affect the operating flexibility for two beams running simultaneously. It would also decrease the efficiency of MT3 for cleaning the beams from backgrounds. This effectively means that the thus modified target area's configuration could only be exploited in runs with a higher program priority for the beamline 24A than for 24B. **Fig. 10** shows the dependences of the antiproton beam intensity and of the $\pi^-$-backgrounds at the experiment target on the position of the production target relative to the upstream edge of the MT3 magnet for the 16 GeV/c beam.

If the center of the additional production target were positioned at +0.7 m into MT3 with its magnetic field near a maximum of 1.9 T, the intensity of antiprotons would increase by a factor of ~2.5 (up to ~$10^6$ per cycle) compared to the main production target position at -0.2 m. But the background from $K_s^0$-decays would grow up even more − by ~3.5 times. The sharp increase of the background from secondary $\pi^-$-mesons produced in the target makes it infeasible moving the target deeper into the MT3 beyond the point of +0.7 m.

## 7. Effects of materials in the beamline 24A

The materials of various beam detectors, as well as the air in the gaps for their installation, affect the characteristics of the beams. An evaluation of the effects of scattering and absorption in materials on the beam properties has been performed under the following assumptions about equipment located along the beamline 24A:

- Three total-flux scintillator counters, 4 mm thick each, located after the quadrupole lenses Q8-Q10 along the beam.
- 13 planes of the beam-tagging scintillator hodoscopes, with the effective thickness of 4 mm each, for beam tagging. Five of them are located at the intermediate focus area in between the quadrupole lenses Q5 & Q6. Four more are positioned at the both ends of the gaps before and after dipole magnets M3 & M4. The remaining four hodoscopes are installed before the experimental target.
- Two threshold Cherenkov counters, 6 m long each, at the stretches through the quadrupoles Q6-Q7 & Q9-Q10. It has been assumed that, for the 45 GeV/c beam, these counters are filled with air at the pressure of 0.6 atm, and for the 15 GeV/c beam, the both are filled with Freon R-22 ($CHClF_2$) at the atmospheric pressure[3].

Particles interactions in Mylar membranes of the thickness 0.5 mm at the ends of vacuum pipes have been also taken into account.

In **Table 4,** the estimations of the effects of scattering and absorption in materials for some beam parameters at the experiment target are shown for '$\delta$-rays', those are the beams of zero transverse size and divergence before entering the beamline. The numbers presented in **Table 4,** as compared to those from **Table 2,** indicate a just minor increase of spatial dimensions for 45 GeV/c beam and about 10-12% increase – for 15 GeV/c beam at the experiment target. The intensity losses due to interactions in the materials are estimated at 14% and 17% for the beams with momenta 45 GeV/c and 15 GeV/c, respectively.

**Table 4.** Parameters of $\delta$-rays at the end of the beamline 24A after passing through the materials.

| $\delta$-ray momentum (GeV/c) | 15 | 45 |
|---|---|---|
| Spatial dimensions $\sigma_x \times \sigma_y$ (mm) | $9.3 \times 7.6$ | $2.7 \times 1.6$ |
| Angular divergences $\sigma_{x'} \times \sigma_{y'}$ (mrad) | $0.53 \times 0.57$ | $0.14 \times 0.24$ |
| Intensity loss (%) | 17 | 14 |

## 8. Conclusion

The new polarized beams for the SPASCHARM experiment will greatly expand and enrich the research program at U-70 accelerator of IHEP in the field of fundamental physics of spin phenomena in hadronic interactions. The design presented in this paper of the beamline 24A provides an opportunity to produce and transport to the experiment polarized protons and antiprotons from $\Lambda(\overline{\Lambda})$-decays at the highest achievable intensities and in the beams of the best quality in the energy range from 10 to 45 GeV. The maximum intensities of proton beams in this energy range vary from ~$10^7$ to ~$10^8$ per accelerator cycle with $10^{13}$ protons on the target, whereas the maximum intensity of the 16 GeV/c antiproton beam could be about $10^6$ per cycle.

The beam optics is optimized for the configuration of two beams operating simultaneously from a single production target and in compliance with the constraints on available space in the U-70 experimental halls, populated with other experiments. The beams of 60 GeV primary protons, of secondary charged particles produced in the production target, as well as electron and/or positron beams will be also available to the SPASCHARM experiment via the beamline 24A.

---

[3] Under these conditions the thresholds for $\pi$-mesons, $K$-mesons and protons are equal to 7.5 GeV/c, 26.4 GeV/c and 50.1 GeV/c in the first case, and 3.5 GeV/c, 12.3 GeV/c and 23.5 GeV/c in the second one. In both cases this provides the suppression of the background of lighter particles, predominantly $\pi$-mesons, in the (anti)proton beam.

**Acknowledgements**

We thank the NRC "Kurchatov Institute" - IHEP management for their support of the beamline 24A and SPASCHARM projects. This work has been supported in part by the Russian Foundation for Basic Research (RFBR) Grant No. 16-02-00667 and by the Competitiveness Programme of National Research Nuclear University MEPhI.